\documentstyle[epsf,eqsecnum,aps]{revtex}

\begin{document}
%\hspace{-15mm}
%\leftline{\epsfbox{mark.eps}}  
%\vspace{-10.0mm} % for revtex
%\vspace{-9.3mm} % for article, article  11pt
%\vspace{-10.3mm} % for article 12pt, revtex preprint style

\thispagestyle{empty}
{\baselineskip-4pt
\font\yitp=cmmib10 scaled\magstep2
\font\elevenmib=cmmib10 scaled\magstep1  \skewchar\elevenmib='177
\leftline{\baselineskip20pt
%\hspace{12mm} % for revtex
%\hspace{15mm} % for article, revtex preprint style
\vbox to0pt
   { {\yitp\hbox{Osaka \hspace{1.5mm} University} }
%\vspace{-4mm} % for revtex preprint style
     {\large\sl\hbox{{Theoretical Astrophysics}} }\vss}}

\rightline{\large\baselineskip20pt\rm\vbox to20pt{
\baselineskip14pt
\hbox{OU-TAP 82}
\vspace{2mm}
\hbox{August 9}\vss}}

\vspace{5mm} 

\begin{center}
{\Large\bf No supercritical supercurvature mode conjecture 
}\\  
{\Large\bf in one-bubble open inflation
}
\end{center}
%\vspace*{8mm}

\centerline{\large Takahiro Tanaka\footnote{Electronic address: 
tama@vega.ess.sci.osaka-u.ac.jp} and Misao Sasaki\footnote{
Electronic address: misao@vega.ess.sci.osaka-u.ac.jp} 
}
\vspace*{2mm}
\begin{center}{\em $^1$Department of Earth and Space Science, 
Graduate School of Science} \\
{\em  Osaka University, Toyonaka 560-0043, Japan}
\end{center}

%\vspace*{8mm}

\begin{abstract}
In the path integral approach to false vacuum decay with the effect of
gravity, there is an unsolved problem, called the negative mode problem.
We show that the appearance of a supercritical 
supercurvature mode in the one-bubble open inflation 
scenario is equivalent to the existence of a negative mode around the
Euclidean bounce solution. Supercritical supercurvature modes are
those whose mode functions diverge exponentially for large spatial
radius on the time constant hypersurface of the open universe.
Then we propose a conjecture that
there should be ``no supercritical supercurvature mode''. 
For a class of models that contains a wide variety of tunneling
potentials, this conjecture is shown to be correct.
\end{abstract}
%\pacs{} 

%\centerline{\bf \today}

\section{introduction}

The Euclidean path integral approach has been used to 
investigate the true vacuum bubble nucleation 
through quantum tunneling\cite{Colema,CalCol}. 
In the lowest WKB approximation, the quantum tunneling 
is described by a bounce solution. 
The bounce solution is a solution of the 
Euclidean equation of motion, 
which connects the configurations before and after tunneling. 
The bounce solution that takes account of the gravitational effect 
was found by Coleman and De Luccia\cite{ColDeL}. 

The decay rate per unit volume and per unit time interval,  
$\Gamma$, is given by the formula\cite{Colema}
\begin{equation}
 \Gamma=|K|e^{-(S_E^{(bounce)}-S_E^{(trivial)})}, 
\end{equation}
where $S_E^{(bounce)}$ is the classical Euclidean action 
for the bounce solution and $S_E^{(trivial)}$ is that 
for the trivial solution that stays at the false vacuum. 
In the path integral approach, 
the prefactor $|K|$ is evaluated by the 
gaussian integral over fluctuations around the 
background bounce solution. 
In a standard system which does not take account of gravity, 
there is one perturbation mode in which direction 
the action decreases. 
It is called a negative mode.   
For this mode, the gaussian integral is not well-defined. 
To make the integral finite, the integration path should be 
deformed on the complex plane. 
Consequently, one imaginary factor, $i$, appears in 
$K$. In the Euclidean path integral approach to 
tunneling, this imaginary unit plays 
a crucial role to interpret $\Gamma$ as the decay rate. 

However, in the case when gravity is taken into account, 
the situation changes drastically.
Since there are gauge degrees of freedom, 
we have various possibilities in 
choosing variables to describe the physical degrees of freedom. 
If we choose inappropriate variables, 
the equation that determines the fluctuation mode can become singular. 
In Ref.\cite{TS92}, 
we have shown that it is impossible to 
obtain a well-behaved reduced action as long as we stick to
variables that appear in the original Lagrangian in the second order
formalism. In order to find variables that can lead us to a well-behaved
reduced action, it was necessary to resort to the Hamiltonian formalism. 
In the Hamiltonian formalism, conjugate momenta are introduced, and 
hence wider varieties of choice of variables are allowed. 
In Ref.\cite{TS92}, we derived a well-behaved reduced action for
fluctuations around the $O(4)$-symmetric bounce solution.  
We found the action for fluctuations which 
conserve the $O(4)$-symmetry has an unusual signature.
Namely the kinetic term is negative definite. 
To deal with this action, we proposed a prescription analogous to the
conformal rotation\cite{GiHaPe}. 
Then, from the path integral measure, there arises one 
imaginary unit $i$. This suggests there should be no negative mode
in the final form of the reduced action, since the factor $i$ in $K$ 
has been already taken care of by the above mentioned prescription. 
Therefore, we proposed the ``no negative mode conjecture''\cite{TS92}. 

On the other hand, in recent years
the process of false vacuum decay with the effect of gravity 
has been studied extensively in the context of 
the one-bubble open inflation scenario. 
In this scenario, an open universe is created inside a nucleated bubble.
In one-bubble open inflation, one of the most important issues is to
calculate the spectrum of quantum fluctuations after the bubble
nucleation because it determines the spectrum of cosmological
perturbations. By comparing the predicted spectrum with the observed one,
we can test a model of one-bubble open inflation. 

{}Fluctuations in an open universe can be decomposed by using spatial
harmonics on the unit 3-dimensional hyperbolic space.
We denote the eigenvalue of a spatial harmonic by $-(p^2+1)$. 
The spatial harmonics with positive $p^2$ are square-integrable 
functions on a time constant hypersurface in an open universe 
in the sense that they can be normalized by using the Dirac 
delta function. As a result the spectrum is continuous for $p^2>0$. 
On the other hand, the spatial harmonics are no longer square-integrable
for $p^2<0$. 
However, since a time constant hypersurface in an open universe 
is not a Cauchy surface, this divergence does not directly 
exclude such modes. By considering the normalization of perturbation
modes on a Cauchy surface, we find that the spectrum for $p^2<0$ becomes
discrete. These modes are called supercurvature modes since they give
rise to correlation on scales greater than the spatial curvature
scale\cite{STY95,LyWo}. 

There are two classes of supercurvature modes, which we call
supercritical and subcritical modes.
The precise definition will be presented later. 
Here we mention that supercritical modes have smaller values (i.e.,
larger absolute values) of $p^2$ than subcritical modes, and their mode
functions diverge exponentially for large spatial radius in the open
universe. We shall show that the existence of a supercritical
supercurvature mode is equivalent to the existence of a negative mode.
Thus the ``no negative mode conjecture'' presented in Ref.\cite{TS92} 
can be restated as the ``no supercritical supercurvature mode
conjecture''. Note that there exists no supercurvature mode for a model
with sufficiently thin bubble wall\cite{GXMT2}, hence there is no
supercritical mode in such a model. 
The purpose of this paper is to examine this conjecture for
a wider class of models. In particular, a class of models we consider
naturally allows thick bubble wall solutions. Recently the negative mode 
problem has been discussed by Lavrelashvili\cite{Lav} in the context of
the singular Hawking-Turok instanton\cite{HaTu}. In this paper, however,
we exclude the possibility of the Hawking-Turok instanton and focus on
regular bounce solutions.

In section 2, we show the equivalence between the existence of a
negative mode in the reduced Euclidean action and that of a
supercritical supercurvature mode in the spectrum of cosmological
perturbations. 
In section 3, we give a method to construct potential models 
which allow an analytical treatment.
By using this method, we give a set of models 
which seemingly violate our conjecture. 
In section 4, 
for such models that have a bounce solution with a negative mode, 
we show that there should be another bounce solution that has
a smaller value of the action and has no negative mode. 
Conclusion is given in section 5.

\section{negative mode and supercritical supercurvature mode}

We consider a system composed of a real scalar field, $\Phi$, 
coupled with the Einstein gravity.  
The Euclidean action is given by 
\begin{equation}
S_E=\int d^4x\,\sqrt{g}\left[
-{1\over 2\kappa}R+{1\over 2} g^{\mu\nu}
\partial_{\mu}\Phi\partial_{\nu}\Phi+V(\Phi)\right]. 
\end{equation}
The potential of the scalar field is assumed 
to have the form as shown in Fig.~\ref{vacdecay}, and initially 
the field is assumed to be trapped in the false vacuum. 
As mentioned in Introduction, 
the bounce solution is a Euclidean solution that connects the
configurations before and after tunneling.  
In the present case, the geometry before false vacuum decay is given by 
a de Sitter space. 
After tunneling, there appears a true vacuum bubble in the false 
vacuum sea. 
This bounce solution is obtained 
by Coleman and De Luccia\cite{ColDeL} 
under the assumption of the $O(4)$-symmetry:
\begin{equation}
 ds^2=a^2(\eta)\left\{d\eta^2+d\chi^2
    +\sin^2\chi\, d\Omega_{(2)}^2\right\},
\quad
 \Phi=\phi(\eta). 
\end{equation}
The Euclidean equations of motion are
\begin{eqnarray}
 && \phi''+2{\cal H}\phi'-a^2 {dV(\phi)\over d\phi}=0,
\label{feq}
\\
 && {\cal H}^2-1={\kappa\over 3}\left({1\over 2}{\phi'}^2
    -a^2V(\phi)\right), 
\label{FRWeq}
\\
 && {\cal H}'-{\cal H}^2 +1 = -{\kappa\over 2}{\phi'}^2, 
\label{calHp}
\end{eqnarray}
where a prime represents the differentiation with respect to $\eta$ 
and ${\cal H}:=a'/a$. 
Requiring the regularity of the bounce solution, 
the boundary condition is determined as 
\begin{equation}
 \phi'\propto e^{-2|\eta|},\quad
 a\propto e^{-|\eta|},\quad (\eta\to \mp\infty). 
\end{equation}
By choosing the initial value of $\phi$ at $\eta\to -\infty$ 
appropriately, we obtain a solution which satisfies the 
required boundary condition. 
For definiteness, we choose $\phi$ to be in the 
false vacuum side for $\eta\to-\infty$.

\vspace{5mm}
\begin{figure}[htb]
\centerline{\epsfbox{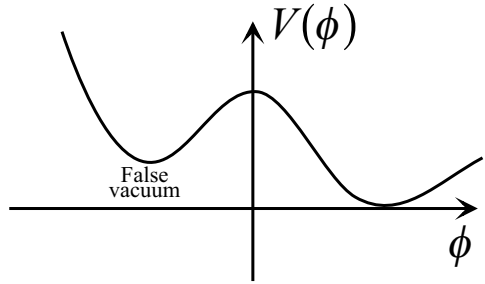}}
\vspace{3mm}
\caption{A typical shape of a scalar field potential under
  consideration.}
\label{vacdecay}
\end{figure}
\vspace{5mm}

As noted before, the number of $i$'s in the prefactor $K$ has a
crucial meaning in the path integral approach. 
To evaluate this number, we need to derive the 
reduced action for fluctuations around the bounce solution.
The fluctuations can be expanded in terms of the spherical harmonics on
the unit 3-sphere, say $Y_{n\ell m}$ with $n=0$, $1$, $2$, $\cdots$,
which satisfy $\bigl[{}^{(3)}\Delta+n(n+2)\bigr]Y_{n\ell m}=0$.
After appropriate gauge fixing, we obtain\cite{TS92,GXMT1} 
\begin{equation}
 \delta^{(2)}S=\sum_{n\ell m} {(n+3)(n-1)\over 2} 
 \int d\eta \left[-i\pi^{n\ell m}{dq^{n\ell m}\over d\eta}
  +{1\over 2}|\pi^{n\ell m}|^2+{1\over 2}(U+(n+3)(n-1))
  |q^{n\ell m}|^2\right], 
\label{action2}
\end{equation}
where $q^{n\ell m}$ is the coefficient of the 
harmonic expansion of a gauge-invariant variable $q$, and
$\pi^{n\ell m}$ its conjugate. In our original derivation\cite{TS92}, we
used the variable $h:=q/a$ and its conjugate momentum. The variable $q$
is equal to the Euclidean version of $\bbox{q}$ introduced
in Ref.\cite{GXMT1}. 
Here, the potential $U$ is given by 
\begin{equation}
 U={\kappa\over 2}\phi'{}^2-W'+W^2\;\quad W={\phi''\over \phi'}\,.
\label{Udef}
\end{equation}
Physically, the variable $q$ represents the curvature perturbation in
the Newton gauge when analytically continued to the open universe inside 
the bubble. The explicit form of the perturbation in this gauge is
\begin{eqnarray}
 ds^2 = && a^2\left[\left(1+{\kappa \phi'q\over a}\right) d\eta^2 
   +\left(1-{\kappa\phi' q\over a}\right) 
    \left(d\chi^2 +\sin^2\chi d\Omega_{(2)}^2 \right)\right], 
\cr 
 \varphi:=&&\Phi-\phi={1\over \phi'}{d\over d\eta}(\phi'q),  
\label{Ngauge}
\end{eqnarray}
where $\varphi$ is the perturbation of the scalar field. 
Recall that the prefactor $K$ is evaluated as\cite{Colema} 
\begin{equation}
K=\int [d\pi\,dq]e^{-\delta^{(2)} S}.
\end{equation}

Looking at Eq.~(\ref{action2}), we find 
that the action disappears for $n=1$. 
This just reflects the fact that the $n=1$ modes are pure gauge and
there is no physical degree of freedom in them.
This is very different from the case of false vacuum decay in the flat
spacetime, in which there is a zero mode in the $n=1$ modes. 
This zero mode describes spacetime translation modes and its existence
implies the existence of a unique negative mode in the $n=0$ (i.e.,
$O(4)$ symmetric) modes\cite{Coluniq}.
In the present case of false vacuum decay with gravity, 
the coefficient in front of $|\pi^{n\ell m}|^2$ 
becomes negative for $n=0$.
Therefore if we try to perform the integration with respect to $\pi$, 
we find that the gaussian integral does not converge for $n=0$. 
To resolve this difficulty, we proposed in \cite{TS92} a prescription
analogous to the conformal rotation.
By changing the variables as $\pi\to -i\pi$, $q\to iq$, 
the above path integral becomes well-defined. To carry out
the path integral, the variables must be discretized.
Then the numbers of $p$ and $q$ integrals will differ by 1. 
Therefore, this change of variables will produce one 
imaginary unit $i$ for the prefactor $K$. 
This is in contrast to the case of false vacuum decay in the flat
spacetime in which the imaginary unit $i$ in $K$ arises from the
negative mode. Note that there is no negative mode for
$n\geq2$\cite{Lav}.

Let us examine the $n=0$ case in more detail. 
If we know the spectrum of eigenvalues, $\lambda_j$, 
of the following Shr\"odinger-type equation, 
\begin{equation}
 \left(-{d^2\over d\eta^2}+U-3\right)q_{j}=\lambda_j q_{j},
\label{yeq}
\end{equation}
where $q_j$ is the $j$-th eigenfunction,  
the contribution to $K$ from $n=0$ modes, 
will be given by $\sim (-i)\prod \lambda_j^{-1/2}$, where the factor
$(-i)$ comes from the rotation of the variables as explained above.
If there is a mode with a negative eigenvalue, 
there arises another imaginary factor, and hence 
the prefactor $K$ becomes real.
If so, this bounce solution will not contribute to the tunneling process
in the path integral approach.
On the other hand, there is another approach to describe the 
tunneling. That is the approach to 
construct a WKB wave function\cite{GerSak,TS92}. 
In this approach, any bounce solution that connects false and true vacua 
will contribute to the tunneling, putting aside the issue of which
bounce dominates.
Thus, in order that the consistency is kept between 
the path integral approach and the wave function approach, 
there should be no negative mode for the Coleman-De Luccia 
bounce solution, provided it gives the smallest Euclidean action among
the non-trivial bounce solutions. 
This is the conjecture proposed in Ref.\cite{TS92}.\footnote{Rigorously
 speaking, there can be even number of negative modes, depending on the
 choice of the canonical variables. However, for the present choice of
 the variables $(\pi,q)$, we conjecture that there is no negative mode.}

\vspace{5mm}
\begin{figure}[htb]
\centerline{\epsfbox{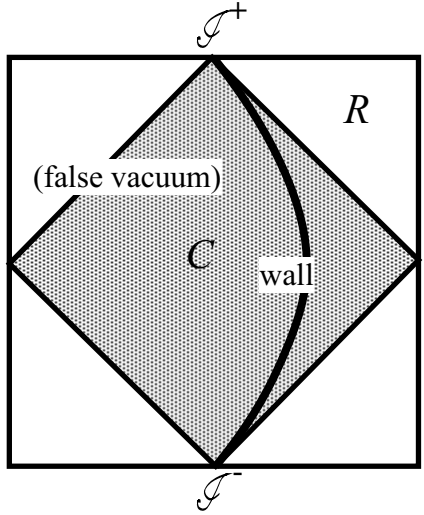}}
\vspace{3mm}
\caption{Conformal diagram of a de Sitter-like space
with a bubble wall. The Cauchy surface exists in region $C$ but
not in the open universe inside bubble, the region marked as $R$.}
\label{deSitt}
\end{figure}
\vspace{5mm}

Now, let us turn to the problem  
of the perturbation spectrum in the context of the 
one-bubble open inflation scenario. 
To study it, we need to 
quantize the perturbation field on the non-trivial 
background that appears after tunneling. 
For this purpose, the reduced action for the perturbation field 
must be calculated. 
In doing so, the time coordinate must be chosen so that 
the time constant hypersurface is a Cauchy surface. 
A convenient choice is to use the coordinate 
that is obtained by the analytic continuation of $\chi$:
$\chi_c=i(\chi-\pi/2)$.
Then, the coordinates $(\chi_c,\eta,\Omega_{(2)})$ span the 
region $C$ of Fig.~\ref{deSitt}. 
In region $C$, the metric takes the form
\begin{equation}
 ds^2=a^2(\eta)
\left(d\eta^2-d\chi_c^2+\cosh^2\chi_c\,d\Omega_{(2)}^2\right).
\end{equation}
It should be remembered that $\eta$ is not a time coordinate there.
The reduced action in this region is obtained as 
\begin{equation}
 S^{(2)}=\sum_{\ell,m}
      {1\over 2}\int d\chi_c\int {d\eta} 
     \left[\cosh^2\chi_c 
       {\partial\overline{{\bbox{q}}^{\ell m}}\over \partial \chi_c}
         \hat{\cal O}{\partial{\bbox{q}}^{\ell m}\over \partial \chi_c}
      -\overline{{\bbox{q}}^{\ell m}}\hat{\cal O} 
      \left\{{\ell(\ell+1)}+(\hat{\cal O}-3)
         \cosh^2\chi_c \right\}{\bbox{q}}^{\ell m}\right], 
\label{actpre}
\end{equation}
where $\hat{\cal O}$ is defined by 
\begin{equation}
 \hat{\cal O}:=-{\partial^2\over \partial\eta^2}+U,
\end{equation}
and $U$ is the same as the one given in Eq.(\ref{Udef}).
In this expression, the $\eta$-dependence 
appears only through the operator $\hat O$. 
Thus the perturbation can be expanded by using eigenfunctions of
$\hat O$: ${\bbox{q}}^{\ell m}
=\sum_p q^p(\eta)f^{p\ell m}(\chi_c)$. 
The equation that determines eigenvalues and eigenfunctions is
\begin{equation}
 \left[-{d^2\over d\eta^2}+U\right] q^{p}
  =(p^2+4) q^{p}. 
\label{qpeq}
\end{equation}
Analyzing the asymptotic behavior of $\phi$, we find $U\to 4$ for
$\eta\to \mp\infty$. Hence the spectrum is continuous for $p^2>0$. 
On the other hand, the spectrum for $p^2<0$ is discrete, which are called
supercurvature modes. Here we call modes with $p^2<-1$ supercritical 
supercurvature, while we call 
modes with $0>p^2>-1$ subcritical supercurvature. When analytically
continued to the open universe by $\chi_R=\chi_c+i\pi/2$, the
supercritical modes diverge exponentially for $\chi_R\to\infty$.
By comparing Eqs.~(\ref{yeq}) and (\ref{qpeq}), 
the existence of a negative mode around the bounce solution is
manifestly equivalent to that of a supercritical supercurvature mode
in the spectrum of quantum fluctuations in the open universe.
Thus the conjecture that we stated above can be 
restated as the ``no supercritical 
supercurvature mode conjecture''. 

\section{model construction} 

At first sight, there seems no reason to deny the possibility of a
potential model that allows a bounce solution with a supercritical
supercurvature mode. In fact, there is a method to construct such
models, as described in a separate paper\cite{GXMT2}. 
However, as mentioned there, this method does not guarantee that 
thus obtained bounce solution gives the smallest Euclidean
action among non-trivial solutions. Hence, given a bounce solution that
allows a supercritical supercurvature mode, if one can prove that its
action is not the smallest, our conjecture remains intact.

In this section, we review the method to construct a potential model and 
a bounce solution, and give an example in which a supercritical
supercurvature mode appears when a model parameter is varied. 
We then construct a potential model that allows a continuous series of
bounce solutions under the weak gravitational backreaction
approximation, which will be used for the investigation of the negative
mode problem in the next section.

Our method for model construction is as follows. 
We first give $\phi'$. It can be specified rather arbitrarily except for 
the conditions that it vanishes as $e^{-2|\eta|}$ for
$\eta\to\mp\infty$ and it is positive (or negative) definite so that
$\phi$ is a monotonic function of $\eta$.
Then we solve the equations (\ref{calHp}) to find ${\cal H}$ and $a$. 
For $\eta\to\mp\infty$, ${\cal H}$ is solved as 
$$
 {{\cal H}-1\over {\cal H}+1}=C_{\mp}e^{2\eta}\,.
$$
Thus the required boundary condition, ${\cal H}\to\pm 1$ for
$\eta\to\mp\infty$, is automatically satisfied. 
Once we know ${\cal H}$, we can determine $a$ by 
integrating ${\cal H}$. Here an integration constant appears, which
determines the absolute value of the potential, say the potential at 
the false vacuum $V_F$. Finally, using Eq.~(\ref{FRWeq}),
the potential $V$ is determined as a function of $\eta$ as
\begin{equation}
 V(\phi(\eta))={6-6{\cal H}^2+{\kappa}{\phi'}^2\over 2\kappa a^2}\,.
\end{equation}
By assumption, since $\phi$ is a monotonic function of $\eta$, by
inversely expressing $\eta$ in terms of $\phi$,
the shape of the potential is uniquely determined.

Next we turn to the eigenvalue equation (\ref{yeq}).
If there is a negative mode, 
the solution with $\lambda=0$ must have, at least, one node.  
Hence, it is sufficient to examine the equation for ${\lambda=0}$, 
which is written as 
\begin{equation}
\left[-{d^2\over d\eta^2}+{\kappa\over 2} {\phi'}^2-W'+W^2-3\right]
q=0. 
\label{sseq}
\end{equation}
If the second term, $\kappa\phi'{}^2/2$ is large enough, the potential 
will stay positive and hence $q$ will have no node. 
Since the change in the overall amplitude of $\phi'$ does not 
alter the function $W$, we can construct a model 
in which the second term is negligibly small without changing the 
other terms. 
Hence, we neglect this positive definite term for the moment.

{}From the boundary condition, we find that $W$ behaves as 
$$
 W\to\pm 2,\quad (\eta\to \mp\infty). 
$$
Except for this boundary condition, $W$ can be arbitrarily chosen. 
Then, by choosing $W$ to stay close to zero for a sufficiently long 
interval of $\eta$, the potential term, $-W'+W^2-3$, 
will have a wide negative-valued region, 
and the solution for $q$ will have a node. 
In contrast, if we consider a sufficiently thin wall, the $W'$ term will 
dominate the potential when $|W|$ is small and $W'<0$. 
Then, the potential will stay positive, and hence $q$ has no
node\cite{GXMT2}.

For instance, let us consider a set of models in which 
$W$ is given by 
\begin{equation}
W=-2\tanh((\eta-\zeta)/\Delta), 
\end{equation}
where the model parameters $\zeta$ and $\Delta$ 
represent the wall location and its thickness, respectively. This is
realized by setting
\begin{equation}
\phi'={\mu\over\left(\cosh((\eta-\zeta)/\Delta)\right)^{2\Delta}}\,,
\end{equation}
where $\mu$ is a constant.
Then, neglecting the $\kappa\phi'{}^2/2$ term, Eq.~(\ref{sseq}) becomes
\begin{equation}
 \left[-{d^2\over d\eta^2}+
\left({2\over\Delta}-4\right)
{1\over\cosh^2((\eta-\zeta)/\Delta)}+1
\right]q=0. 
\label{Uex}
\end{equation}
For $\Delta=1$, this equation has a regular nodeless solution, 
\begin{equation}
 q={1\over \cosh(\eta-\zeta)}\,.
\label{critsol}
\end{equation}
Thus the lowest eigenvalue is zero for $\Delta=1$.
Since the potential term in Eq.~(\ref{Uex}) is a monotonically 
decreasing function of $\Delta$, 
there will be a negative mode for $\Delta>1$.
Now let us turn on the $\kappa\phi'{}^2/2$ term while keeping $W$ the
same. As we increase $\mu$, the critical value of $\Delta$ for the
existence of a negative mode will increase. For $\kappa\mu^2\gtrsim1$, a
negative mode will disappear completely for any value of $\Delta$.

To make the problem tractable, let us consider the weak backreaction 
limit, in which the geometry can be approximated by 
a pure de Sitter space. 
{}From Eq.~(\ref{calHp}), we see that this will be the case 
if $\kappa\phi'{}^2\ll 1$.
In the above model, we have seen that a negative mode will disappear if
the $\kappa\phi'{}^2/2$ term gives a large contribution (of order unity
or greater) to the potential. This will be true for most conceivable
potential models. Hence we may restrict our attention to a class of
models that automatically satisfy the weak backreaction condition when
investigating the existence of a negative mode.

%To see this, we rewrite Eq.~(\ref{feq}) as
%\begin{equation}
% {d\over d\eta}\left(V
%   -{\kappa\over 2}{\phi'{}^2\over a^2}\right)
%=3{{\cal H}\over a^2} \phi'{}^2.
%\end{equation}
%Integrating this equation, we have
%\begin{equation}
% \kappa a^2(V-V_F)={1\over 2}\kappa\phi'{}^2+3\kappa a^2
% \int {da\over a^3}\phi'{}^2
%\label{energyeq}
%\end{equation}
%In order that $W$ stays small for a sufficient long interval,
%$\kappa \phi'{}^2$ must stay approximately constant.
%Thus we assume $\kappa \phi'{}^2\approx A=$ constant from
%$a=a_i$ to $a=a_f$. Then we will have
%\begin{equation}
% \kappa (V|_{a=a_f}-V_F)\sim{3\kappa A\over 2a_f^2}
%\left(1-{a_f^2\over a_i^2}\right) ,
%\end{equation}
%Since the interval in which $\phi'{}^2\approx A$
%must continue for a sufficiently long inteval,
%we should have $a_f^2/a_i^2\ll 1$.
%Furthermore, we should have $a_f^2<a_i^2<a_{max}^2
%\approx 3/\kappa V_F$, where $a_{max}$ is the maximum value
%of $a$.
%By also using the fact that $V$ must be positive,
%we finally obtain $A\ll 1$ unless $a_f^2\approx a_i^2
%\approx a_{max}^2$. Since the second term
%of the right hand side of Eq.~(\ref{energyeq}),
%we can also say that
%\begin{equation}
%(V_{top}-V_F)/V_F \alt O(\kappa\phi'{}^2),
%\label{Vbound}
%\end{equation}
%where $V_{top}$ is the value of $V$ at the top of the potential barrier.
%Henceforce, we restrict our attention to the
%weak backreaction limit below.
%This restriction simplifies the analysis very much because
%in the weak backreaction limit the problem reduces to
%that on a fixed de Sitter background.

In the first non-trivial order of the weak backreaction approximation,
we find that the negative mode problem can be analyzed on a fixed de
Sitter background geometry. Let us explicitly show this fact.

First we examine the background field equations.
Neglecting relative errors of $O(\kappa\phi'{}^2)$, 
Eq.~(\ref{calHp}) can be readily solved as 
\begin{equation}
 a_b={1\over H\cosh \eta}\,, 
\quad {\cal H}_b=-\tanh\eta\,, 
\end{equation}
where $H$ is defined by $H^2:=(\kappa/3) V_{top}$ where $V_{top}$ is the
value of $V$ at the top of the potential barrier. 
It is worthwhile to note that the weak backreaction approximation does
not imply small potential difference between the false and true vacua.
The vacuum energy inside the bubble can be much smaller than $V_F$ as
long as the bubble radius is sufficiently smaller than $H^{-1}$. 
However, the weak backreaction approximation does imply 
$V_{top}-V_F\ll V_F$. Hence we may replace $V_{top}$ with $V_{F}$ in the
definition of $H$. Here we have chosen $V_{top}$ for later convenience.
Then substituting the lowest order solutions of $a$ and ${\cal H}$ to
Eq.~(\ref{feq}), the scalar field equation becomes 
\begin{equation}
 \phi''+2{\cal H}_b\phi' -a_b^2 {d\over d\phi}(\delta V)=0,
\label{phieq0}
\end{equation}
where $\delta V=V-V_{top}$. This gives the lowest order solution of
$\phi$. Introducing the variable $N:=\log a$, and expanding it as 
$N_b+\delta N$, Eq.~(\ref{calHp}) gives the equation for the first order 
correction to $a$:
\begin{equation}
{1\over \cosh^2\eta}{d\over d\eta}
\left(\cosh^2\eta\,{d(\delta N)\over d\eta}\right)
 =-{\kappa\over 2}\phi'{}^2. 
\end{equation}
Since $\phi'{}^2\sim e^{-4|\eta|}$ for $\eta\to \pm\infty$, 
the above equation can be integrated and we find 
$\delta N=O(\kappa\phi'{}^2)$ everywhere.
If we substitute this $\delta N$ back into Eq.~(\ref{feq}), 
we can calculate the correction of $O(\kappa\phi'{}^2)$ to $\phi$. 

Next we evaluate the action under the weak backreaction approximation.
The Euclidean action for an $O(4)$-symmetric configuration 
is given by 
\begin{eqnarray}
 S_E =&& S_g[a]+S_m[a,\phi];
\nonumber\\
&&S_g[a]:=2\pi^2\int d\eta\left[-{3\over \kappa} a'{}^2 
-{3\over \kappa} a^2 +a^4 V_{top}\right],
\label{Sgdef}\\
&&S_{m}[a,\phi]:=
2\pi^2\int d\eta\left[{a^2\over 2}\phi'{}^2 
 +a^4 \delta V(\phi)\right]. 
\label{Smdef}
\end{eqnarray}
Inserting $a_b+\delta a$ to the gravitational part of the action, where
$\delta a=a_b\delta N=O(\kappa\phi'{}^2)$, we find
\begin{equation}
  S_g[a_b+\delta a]
=S_g[a_b]\left(1+O\left((\kappa\phi'{}^2)^2\right)\right), 
\end{equation}
which follows from the fact that $a_b$ is a solution 
to the equation obtained from the action $S_{g}[a]$. 
Hence, to the first order of $\kappa\phi'{}^2$, we have 
\begin{equation}
S_E=S_g[a_b]+S_m[a_b,\phi],
\end{equation}
where $S_m[a_b,\phi]$ is $O(\kappa\phi'{}^2)$ relative to $S_g[a_b]$.
Thus the action evaluated on a fixed de Sitter background 
is correct to the first non-trivial order. Since the background geometry 
is fixed, we focus on the matter part of the action and denote it by
$S_m[\phi]$ in the following. 

Finally, we consider $O(4)$-symmetric fluctuations around the bounce
solution. We rewrite the eigenvalue equation (\ref{yeq}) by using the
variable $\varphi$ introduced in (\ref{Ngauge}). We find
\begin{eqnarray}
 \varphi''+&&\left(2{\cal H}-{\kappa\phi'\phi''\over {\kappa\over 2} 
     \phi'{}^2 -\lambda-3}\right) \varphi' 
\cr&& + 
 \left(\lambda-2\kappa\phi'{}^2-a^2\partial^2V
    +{\kappa\phi'\phi''\over {\kappa\over 2}\phi'{}^2 -\lambda-3}
    \right) \varphi=0,
\label{phieq}
\end{eqnarray}
where $\partial^2V=d^2V(\phi)/d\phi^2$.
Unless $\lambda$ is close to $-3$, this equation reduces in the weak
backreaction limit to 
\begin{equation}
 \varphi''+2{\cal H}_b\varphi'-a_b^2\partial^2(\delta V)\varphi
=-\lambda \varphi,
\label{varphieq}
\end{equation}
Here we note that the term $a_b^2\partial^2 (\delta V)$ can be as large
as ${\cal H}_b^2$ even in the weak backreaction limit. 
The above equation is the same as the equation for $O(4)$-symmetric
scalar field fluctuations on a fixed de Sitter background, i.e., without
taking account of the metric perturbation.

In the above, we have shown that it is sufficient 
to consider the problem on a fixed de Sitter background 
in the first non-trivial order of the weak backreaction approximation.
Now let us return to the model with $W=-2\tanh(\eta-\zeta)$. 
We assume $\zeta>0$ since the 3-sphere of the 
maximum radius should be on the false vacuum side in order for the
bounce solution to describe false vacuum decay.
As mentioned there, $\Delta=1$ is the critical case in this model
in the weak backreaction limit.
This was shown by the fact that there is a regular, nodeless solution of
Eq.~(\ref{yeq}) with $\lambda=0$ for $\Delta=1$, which is given by
Eq.~(\ref{critsol}), and that the potential $U$ is a monotonically
decreasing function of $\Delta$.
The solution for $\varphi$ that corresponds to the nodeless solution
(\ref{critsol}) for $q$ is given by
\begin{equation}
\varphi\propto {\sinh(\eta-\zeta)\over\cosh^2(\eta-\zeta)}\,.
\label{varphinode}
\end{equation}
This solution has a node. In fact, Eq.~(\ref{varphieq}) has a solution 
$\varphi=\phi'/a_b$ with a lower (hence negative) eigenvalue
$\lambda=-3$. If the gravitational effect were completely neglected,
this would be the negative mode that gives rise to the factor $i$ of the
path integral. However, as we have noted, the approximate equation 
(\ref{varphieq}) is no longer valid for $\lambda\approx -3$ even in the 
weak backreaction limit. Since the unapproximated equation (\ref{phieq})
for $\varphi$ is not a Schr\"odinger-type equation, the correspondence
between the number of nodes and the increasing order of eigenvalues is
lost. Thus the existence of the solution (\ref{varphinode}) with a node
at $\lambda=0$ does not imply the existence of a negative mode.

For $\Delta=1$, the background solution for $\phi$ is given by 
\begin{equation}
 \phi=\phi_{\zeta}:=\phi_* (c_1\tanh(\eta-\zeta)+c_2), 
\end{equation}
where $\phi_*$ is a constant which determines the typical scale of
$\phi$ and $c_1$ and $c_2$ are integration constants to be
determined later. The potential $\delta V$ is reconstructed as 
\begin{eqnarray}
 \delta\tilde V & := & {\delta V\over H^2\phi_*^2}
= {1\over H^2\phi_*^2}\int d\eta \,\partial(\delta V)
 \phi'_{\zeta}
\cr 
& = & {1\over H^2\phi_*^2}
 \int d\eta {\phi'_{\zeta}{}^2\over a_b^2}
 \left(W+2{\cal H}_b\right)
\cr
&= &c_1^2 \int d\eta {\cosh^2\eta \over \cosh^4(\eta-\zeta)}
 \left(-2\tanh(\eta-\zeta)-2\tanh\eta\right)
\cr
&=&-c_1^2{\sinh^2(2\eta-\zeta)\over\cosh^4(\eta-\zeta)}
\cr
 & = & -{c_1^2\over 2\sinh^2\zeta}\left[
\left(\sinh\zeta\,\tanh(\eta-\zeta)+\cosh\zeta\right)^2
-1\right]^2,
\end{eqnarray}
where we have set the integration constant so that 
$\delta V=0$ at the top of the potential.
Then by choosing $c_1$ and $c_2$ as 
\begin{equation}
 c_1={1\over 2}\sinh\zeta,\quad 
 c_2={\cosh\zeta +1\over 2}, 
\end{equation}
we find that all the solutions parametrized by $\zeta$ 
are solutions for the same potential given by\footnote{Jaume Garriga
  pointed out to us that this series of instantons can be obtained by a
  conformal transformation of the Fubini instanton in flat
  space\cite{Fubini}.
  Some implications of these instantons to open inflation
  models are discussed in a separate paper\cite{GXMT2}.}
\begin{equation}
 \delta V=-{2H^2\over\phi_*^2}\phi^2(\phi-\phi_*)^2. 
\end{equation}
Note that the eigenfunction for the $\lambda=0$ mode is
also given by taking the derivative of the background solution 
with respect to $\zeta$; 
$\varphi\propto d\phi_{\zeta}/d\zeta \propto 
\sinh(\eta-\zeta)/\cosh^2(\eta-\zeta)$.
For $\zeta=0$, the scalar field stays still on top of the potential
barrier. This is a Hawking-Moss type solution\cite{HaMo}.
It is easy to see that the action $S_m[\phi_{\zeta=0}]$ vanishes. 
Since $\phi_{\zeta}$ is a continuous series
of solutions, the action must be the same for all the values 
of $\zeta$. This implies $S_m[\phi_{\zeta}]=0$ for any 
$\zeta$. This can be also checked by direct substitution. 

\section{perturbation analysis}
Let us denote the matter part of the action, $S_m[\phi]$, 
corresponding to the critical case of the model discussed in the
previous section by 
\begin{equation}
S^{(0)}[\phi]:=\int d\eta \left({a_b^2\over 2}\phi'{}^2 
  -a_b^4{2H^2\over\phi_*^2}\phi^2(\phi-\phi_*)^2\right). 
\end{equation}
We consider a set of models whose action is given by 
\begin{equation}
 S_m[\phi]=S^{(0)}[\phi]+S^{(1)}[\phi], 
\end{equation}
where $S^{(1)}[\phi]$ is a small perturbation of order $\epsilon\ll1$.
We need not specify the explicit form of $S^{(1)}[\phi]$. 
In the following, we neglect the $O(\epsilon^2)$ terms. 

For convenience, we introduce the following notation for variations of
the action: 
\begin{eqnarray}
 && {\delta S\over \delta\phi}[\phi;\psi]
:=\int d\eta \left.{\delta S\over \delta\phi(\eta)}
\right\vert_{\phi}\psi(\eta),
\cr
 && {\delta S\over \delta\phi}[\phi;*]
:=\left.{\delta S\over \delta\phi(\eta)}
\right\vert_{\phi}\,,
\cr
 && {\delta^2 S\over \delta\phi^2}[\phi;\psi_1,\psi_2] 
  :=\int d\eta_1 d\eta_2 
  \left.{\delta^2 S\over \delta\phi(\eta_1)\delta\phi(\eta_2)}
  \right\vert_{\phi}
  \psi_1(\eta_1)\psi_2(\eta_2),
\cr
&& {\delta^2 S\over \delta\phi^2}[\phi;*,\psi]
  :=\int d\eta'
  \left.{\delta^2 S\over \delta\phi(\eta)\delta\phi(\eta')}
  \right\vert_{\phi} \psi(\eta'),
\end{eqnarray}
and analogous expressions for higher variations.

Let us first write down useful formulas for our discussion below.
Since $\phi_\zeta$ is a solution to the unperturbed equation of motion 
for any value of $\zeta$, successive differentiations of the equation of 
motion with respect to $\zeta$ give
\begin{eqnarray}
&& {\delta S^{(0)}\over \delta\phi}
\left[\vphantom{{d\phi_{\zeta}\over d\zeta}}\phi_{\zeta};*\right]=0, 
\\
&&{\delta^2 S^{(0)}\over \delta\phi^2}\left[\phi_{\zeta};*,
 {d\phi_{\zeta}\over d\zeta}\right]=0, 
\label{BG1}
\\
&&{\delta^3 S^{(0)}\over \delta\phi^3}\left[\phi_{\zeta};*,
 {d\phi_{\zeta}\over d\zeta},{d\phi_{\zeta}\over d\zeta}\right]
 +{\delta^2 S^{(0)}\over \delta\phi^2}\left[\phi_{\zeta};*,
 {d^2\phi_{\zeta}\over d\zeta^2}\right]=0, 
\label{BG2}
\end{eqnarray}
In addition, we have $S^{(0)}[\phi_{\zeta}]=0$ as mentioned at the
end of the previous section. Hence, for arbitrary $\delta\phi$ of
$O(\epsilon)$ we have 
\begin{equation}
 S[\phi_{\zeta}+\delta\phi]
 ={\delta S^{(0)}\over \delta\phi}[\phi_{\zeta};\delta\phi]
 +S^{(1)}[\phi_{\zeta}]
 =S^{(1)}[\phi_{\zeta}]. 
\end{equation}
Then the derivatives of $S[\phi_{\zeta}+\delta\phi]$ with respect 
to $\zeta$ are expressed as
\begin{eqnarray}
&& {d S\over d\zeta}=
  {\delta S^{(1)}\over \delta\phi}\left[\phi_{\zeta};
 {d\phi_{\zeta}\over d \zeta}\right], 
\label{dSdzeta}\\
&&  {d^2 S\over d\zeta^2}=
 {\delta^2 S^{(1)}\over \delta\phi^2}\left[\phi_{\zeta};
 {d\phi_{\zeta}\over d \zeta},
 {d\phi_{\zeta}\over d \zeta}\right]
 + {\delta S^{(1)}\over \delta\phi}\left[\phi_{\zeta};
 {d^2\phi_{\zeta}\over d\zeta^2}\right]. 
\label{dSdzeta2}
\end{eqnarray}

Now we analyze the behavior of the action under a variation of $\zeta$
and its relation to the value of the lowest eigenvalue for
the $O(4)$-symmetric fluctuations.

First we show that $dS[\phi_\zeta+\delta\phi]/d\zeta=0$ if
$\phi_\zeta+\delta\phi$ is a solution to the equation of motion to
$O(\epsilon)$. Note that a bounce solution generally cease to exist for
continuous values of $\zeta$, but it exists only at discrete values of
$\zeta$. Assuming that $\phi_\zeta+\delta\phi$ is a solution, we have
\begin{equation}
 0={\delta S\over \delta\phi}[\phi_{\zeta}+\delta\phi;*]
 ={\delta^2 S^{(0)}\over \delta\phi^2}\left[\phi_{\zeta};*,
    \delta\phi\right]
   +{\delta S^{(1)}\over \delta\phi}[\phi_{\zeta};*].
\label{deltaphi}
\end{equation}
Then $dS/d\zeta$ in Eq.~(\ref{dSdzeta}) can be expressed by using
Eq.~(\ref{deltaphi}) as 
\begin{equation}
 {dS\over d\zeta}=-
 {\delta^2 S^{(0)}\over \delta\phi^2}[\phi_{\zeta};
 {d\phi_{\zeta}\over d \zeta},\delta\phi], 
\end{equation}
and this vanishes because of Eq.~(\ref{BG1}). 

Next we consider the shift of the lowest eigenvalue, $\lambda$, due to
$S^{(1)}$, and show that $d^2 S/d\zeta^2$ has the same signature 
as $\lambda$ does. In the unperturbed case, the lowest eigenvalue is
zero, $\lambda=\lambda^{(0)}=0$, whose eigenfunction is given by
$\varphi^{(0)}=d\phi_\zeta/d\zeta$. 
To evaluate the shift of $\lambda$, 
we consider the eigenvalue problem given by 
\begin{equation}
 {\delta^2 S\over \delta\phi^2}[\phi_{\zeta}+\delta\phi;*,
 \varphi]=\lambda\,a_b^2\,\varphi. 
\label{modeeq}
\end{equation}
In particular, we have 
\begin{equation}
 {\delta^2 S\over \delta\phi^2}
 [\phi_{\zeta}+\delta\phi;\varphi,\varphi]
 =\lambda\int a_b^2\,|\varphi|^2 d\eta. 
\label{varphi2}
\end{equation}
We denote the $O(\epsilon)$ correction to $\varphi$ by $\varphi^{(1)}$:
$\varphi=\varphi^{(0)}+\varphi^{(1)}$.
Using Eq.~(\ref{BG1}), the left hand side of Eq.~(\ref{varphi2})
reduces to
\begin{equation}
  {\delta^2 S\over \delta\phi^2}[\phi_{\zeta}+\delta\phi, 
 \varphi^{(0)},\varphi^{(0)}]
= {\delta^3 S^{(0)}\over \delta\phi^3}[\phi_{\zeta};\delta\phi, 
 \varphi^{(0)},\varphi^{(0)}]
+ {\delta^2 S^{(1)}\over \delta\phi^2}[\phi_{\zeta}, 
 \varphi^{(0)},\varphi^{(0)}]. 
\end{equation}
The first term in the right hand side of the above equation can be
rewritten by using Eqs.~(\ref{BG2}) and (\ref{deltaphi}) as
\begin{equation}
 {\delta^3 S^{(0)}\over \delta\phi^3}[\phi_{\zeta};\delta\phi, 
 \varphi^{(0)},\varphi^{(0)}]
 =-{\delta^2 S^{(0)}\over \delta\phi^2}\left[\phi_{\zeta};\delta\phi, 
 {d\varphi^{(0)}\over d\zeta}\right]
 ={\delta S^{(1)}\over \delta\phi}\left[\phi_{\zeta};
   {d\varphi^{(0)}\over d\zeta}\right]. 
\end{equation}
Thus, to $O(\epsilon)$, we obtain
\begin{equation}
\lambda\int a_b^2|\varphi|^2 d\eta
={\delta^2 S^{(1)}\over \delta\phi^2}[\phi_{\zeta}, 
 \varphi^{(0)},\varphi^{(0)}]
+{\delta S^{(1)}\over \delta\phi}\left[\phi_{\zeta};
   {d\varphi^{(0)}\over d\zeta}\right]={d^2S\over d\zeta^2}\,,
\end{equation}
where the last equality follows from Eq.~(\ref{dSdzeta2}).
Therefore the sign of the eigenvalue $\lambda$ is the same as that of 
$d^2S/d\zeta^2$.
This tells us that when there appears a negative 
mode, i.e., if $\lambda<0$, the action of the bounce solution is not a
local minimum.

To gain a more definite picture of the behavior of the action,
we consider possible models for $S^{(1)}$ with a certain variable
parameter. Among all the possibilities, we give schematic plots of three
typical cases of $S[\phi_{\zeta}+\delta\phi]=S^{(1)}[\phi_\zeta]$ in
Fig.~\ref{Sone}(a) -- (c). 
In each plot, the dotted line represents the case of a parameter for
which the bounce solution has a positive $d^2S/d\zeta^2$. 
As the parameter is varied, the shape of $S^{(1)}[\phi_{\zeta}]$ 
smoothly passes through the one like the dashed line and changes to the
one given by the rigid line. 
In the case (a), the value of $\zeta$ at which $S^{(1)}[\phi_{\zeta}]$
takes an extremum value shifts toward zero,
and only the Hawking-Moss type bounce at $\zeta=0$ is left in the end. 
In the case (b), the value of $\zeta$ at the extremum of
$S^{(1)}[\phi_\zeta]$ does not shift much, but the amplitude of
$S^{(1)}[\phi_\zeta]$ changes. At the moment when
$S^{(1)}[\phi_{\zeta}]$ is given by the dashed line, 
infinite number of solutions become degenerate, just like the
case of the unperturbed action $S^{(0)}$. 
Beyond this critical point, there appears again one non-trivial
solution, but with negative $d^2S/d\zeta^2$. The action 
is manifestly greater than that of the Hawking-Moss solution at
$\zeta=0$. In the case (c), at the moment when $S^{(1)}[\phi_{\zeta}]$ is 
given by the dashed line, there appears a $\lambda=0$ mode at the
extremum of $S^{(1)}[\phi_\zeta]$. Beyond this critical point, there
appear three non-trivial extrema. Two of them that newly appeared
have positive values of $d^2S/d\zeta^2$ while the one corresponding to
the original solution has negative $d^2S/d\zeta^2$, hence has 
a negative mode. Also in this case, the latter solution with a negative
mode has a larger action than that for either of the other two solutions.

\vspace{5mm}
\begin{figure}[htb]
\centerline{\epsfbox{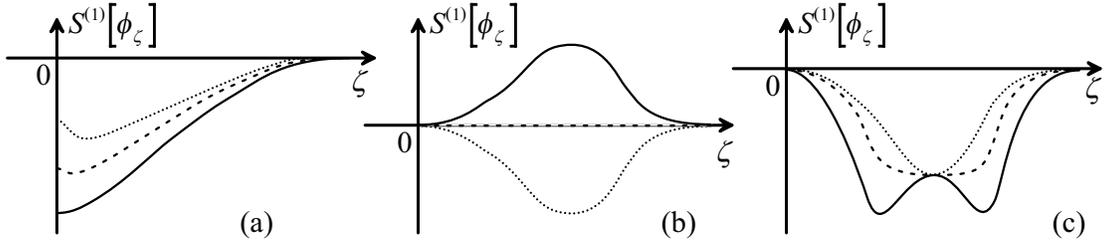}}
\vspace{3mm}
\caption{Three typical cases of the behavior of the Euclidean action
as a function of $\zeta$ when a model parameter is varied.
}
\label{Sone}
\end{figure}
\vspace{5mm}

As an example of $S^{(1)}$, let us consider the one given by 
\begin{equation}
 S^{(1)}=\epsilon\int d\eta\, a_b^4\, V_1,
\quad
 V_1=H^4\sum_{n=0}^4 \alpha_n \,\tilde\phi^n, 
\end{equation}
where $\tilde\phi:=2(\phi/\phi_*)-1$. In this case,
$S^{(1)}[\phi_{\zeta}]$ can be calculated analytically as 
\begin{equation}
 S^{(1)}[\phi_{\zeta}]=\epsilon\left[{4\over 3}(\alpha_0+\alpha_4)
    +{2(\alpha_1+\alpha_3)\over \sinh^3\zeta}
     \left(\cosh\zeta\sinh\zeta-\zeta\right)
    +{4\alpha_2\over \sinh\zeta}
     \left(\zeta\coth\zeta-1 \right)\right]. 
\end{equation}
If we choose a one-parameter family of models, say, like 
$2(\alpha_1+\alpha_3)=-2+\alpha$ and $4\alpha_2=3$, where $\alpha$ is
the variable parameter, the situation as shown in Fig.~\ref{Sone}(a) is
realized. On the other hand, if we choose $2(\alpha_1+\alpha_3)=-\alpha$
and $4\alpha_2=2\alpha$, the situation as shown in Fig.~\ref{Sone}(b) is
realized.  Within this limited class of perturbations, 
the situation as shown in Fig.~\ref{Sone}(c) is not realized. 

\section{conclusion}
We have investigated the negative mode problem associated with false 
vacuum decay with gravity. We have shown that the existence of a
negative mode around a non-trivial Euclidean solution, called the
Coleman-De Luccia bounce solution, is equivalent to that of a
supercritical supercurvature mode in the perturbation spectrum of the
quantum fluctuations in the open universe that appears inside the
bubble. Supercritical supercurvature modes are those for which the mode
functions diverge exponentially for large spatial radius in the open
universe. Then we have proposed a conjecture that there exists no
supercritical supercurvature mode. If this is true, there will be no
negative mode around the Coleman-De Luccia bounce solution that
dominates the process of false vacuum decay.

To investigate the validity of our conjecture, we have first provided a
potential model that admits a continuous series of Coleman-De Luccia
type bounce solutions under the weak gravitational backreaction
approximation. This series contains a Hawking-Moss type solution as a
limiting case and each of these bounce solutions has a zero mode as the
lowest eigenvalue. Then for a class of potentials that can be realized
by small modifications of this potential model,
we have analyzed the behavior of the Euclidean action around a bounce
solution. We have shown that its Euclidean action is not the smallest
among non-trivial solutions if there exists a negative mode.

We have considered three typical cases of the behavior of the action
when a model parameter is varied. In all of these cases, we have found
that, when there appears a bounce solution with a negative mode, it does
not give the smallest action but there exists another bounce
solution, either Coleman-De Luccia type or Hawking-Moss type, with a
lower action that has no negative mode. This evidence strongly supports
the conjecture that there is no negative mode for the Coleman-De Luccia
bounce solution that dominates the tunneling process.

\vspace{1cm}
\centerline{\bf Acknowledgments}
We thank J. Garriga for helpful discussions. 
This work was supported in part by the Monbusho Grant-in-Aid for
Scientific Research No.~09640355 and by the Saneyoshi foundation.

\end{document}